\input harvmac

\rightline{\vbox{\hbox{CERN-TH/97-277}\hbox{\tt hep-th/9710126}}}
\Title{}{Global Anomalies in $M$-theory}
\centerline{M{\aa}ns Henningson}
{\it \centerline{Theory Division, CERN}
\centerline{CH-1211 Geneva 23, Switzerland} \centerline{\tt
henning@nxth04.cern.ch}}

\bigskip \bigskip \bigskip \centerline{\bf Abstract}
We first consider $M$-theory formulated on an open eleven-dimensional
spin-manifold. There is then a potential anomaly under gauge
transformations on the $E_8$ bundle that is defined over the boundary
and also under diffeomorphisms of the boundary. We then consider
$M$-theory configurations that include a five-brane. In this case,
diffeomorphisms of the eleven-manifold induce diffeomorphisms of the
five-brane world-volume and gauge transformations on its normal
bundle. These transformations are also potentially anomalous. In both
of these cases, it has previously been shown that the perturbative
anomalies, i.e. the anomalies under transformations that can be
continuously connected to the identity, cancel. We extend this
analysis to global anomalies, i.e. anomalies under transformations in
other components of the group of gauge transformations and
diffeomorphisms. These anomalies are given by certain topological
invariants, that we explicitly construct.

\Date{\vbox{\line{CERN-TH/97-277\hfill} \line{October 1997 \hfill}}}

\newsec{Introduction}
The consistency of a theory with gauge-fields or dynamical gravity
requires that the effective action is invariant under gauge
transformations and space-time diffeomorphisms, usually referred to
as cancelation of gauge and gravitational anomalies. The first step
towards establishing that a given theory is anomaly free is to
consider transformations that are continuously connected to the
identity. The cancelation of the corresponding anomalies, often
called perturbative anomalies, imposes some constraints on the chiral
field content of the theory. Given that the perturbative anomalies
cancel, it makes sense to investigate transformations in other
components of the group of gauge transformations and diffeomorphisms.
An anomaly under such a transformation is usually referred to as a
global anomaly. A general formula for the quantum contribution of
chiral fields to global anomalies was given in \ref\Wittenone{
E. Witten, `Global gravitational anomalies', {\it Commun. Math.
Phys.} {\bf 100} (1985) 197.
}. Provided that the perturbative anomalies cancel, the global
anomaly is a topological invariant, i.e. it it invariant under smooth
deformations of the data, und thus only depends on the topological
classes of for example the space-time manifold and the gauge-bundle.
The requirement that the anomaly vanishes for an arbitrary
transformation imposes some restrictions on these objects.

In string theory, the requirements of supersymmetry and vanishing
anomalies are particularly constraining, because of the high
dimensionality of space-time. For the purpose of computing anomalies,
it is enough to know the low-energy effective supergravity theory.
The non-chiral type IIA supergravity theory obviously has no
anomalies. The chiral type IIB supergravity theory has a potential
perturbative gravitational anomaly, but the contributions from the
various chiral fields `miraculously' cancel against each other
\ref\AlvarezGaumeWitten{
L. Alvarez-Gaum\'e and E. Witten, `Gravitational anomalies', {\it
Nucl. Phys.} {\bf B234} (1983) 269.
}. Type I supergravity coupled to some super-Yang-Mills theory always
has a non-vanishing perturbative quantum anomaly. It can however be
cancelled by a Green-Schwarz mechanism \ref\GreenSchwarz{
M. B. Green and J. H. Schwarz, `Anomaly cancellations in $D = 10$
gauge theory and superstring theory', {\it Phys. Lett.} {\bf 149B}
(1984) 117.
} involving an anomalous transformation law at tree-level for the
two-form field, provided that the gauge group is $SO(32)$ or $E_8
\times E_8$. This discovery actually preceeded the construction of
the $SO(32)$ and $E_8 \times E_8$ heterotic string theories. Global
anomalies in string theory was first considered in \Wittenone. The
result is that the known string theories are free from global
anomalies when formulated in ten-dimensional Minkowski space.
However, there are interesting non-trivial restrictions on consistent
compactifictions to lower dimensions.

There is by now mounting evidence that the different string theories
should be seen as particular limits of a conjectured theory called
$M$-theory. In the long wave-length limit, this theory should reduce
to eleven-dimensional supergravity. It should also admit certain
types of topological defects, in particular space-time boundaries and
five-brane solitons. Much has been learned about $M$-theory by
studying the mechanism for cancellation of perturbative anomalies for
various configurations including such defects. In this way, it was
discovered that when $M$-theory is defined on an open
eleven-manifold, there is an $E_8 \times E_8$ super-Yang-Mills
multiplet propagating on the boundary. The cancellation of
perturbative gauge and gravitational anomalies involves a subtle
interplay between contributions from this multiplet, the bulk degrees
of freedom and a generalized Green-Schwarz mechanism
\ref\HoravaWitten{
P. Horava and E. Witten, `Heterotic and type I string dynamics from
eleven dimensions', {\it Nucl. Phys.} {\bf B460} (1996) 506, {\tt
hep-th/9510209} \semi
P. Horava and E. Witten, `Eleven dimensional supergravity on a
manifold with boundary', {\it Nucl. Phys.} {\bf B475} (1996) 94, {\tt
hep-th/9603142}.
}\ref\deAlwisone{
S. P. de Alwis, `Anomaly cancellation in $M$-theory', {\it Phys.
Lett.} {\bf B392} (1997) 332, {\tt hep-th/9609211}.
}. The situation is even more interesting for $M$-theory
configurations involving a five-brane. The perturbative gauge and
gravitational anomalies from the five-brane world-volume theory can
be partially cancelled by an anomaly inflow from the surrounding
eleven-dimensional space \ref\DuffLiuMinasian{
M. Duff, J. Liu and R. Minasian, `Eleven-dimensional origin of
string/string duality: a one loop test', {\it Nucl. Phys.} {\bf B452}
(1995) 261, {\tt hep-th/9506126}.
}\ref\Wittentwo{
E. Witten, `Five-brane effective action in $M$-theory', {\tt
hep-th/9610234}.
}. There is however a remaining part, whose cancellation seems to
require additional world-volume interactions and also imposes a
certain topological restriction on the five-brane configuration
\Wittentwo\ref\deAlwistwo{
S. P. de Alwis, `Couplings of branes and normalization of effective
actions in string/$M$-theory', {\tt hep-th/9705139}.
}.

Given our present incomplete understanding of $M$-theory, it seems
that any further information about this theory would be valuable. The
purpose of the present paper is to carry the analysis described above
one step further by investigating also global anomalies. In section
two, we consider the case of $M$-theory defined on an open
eleven-manifold. In section three, we instead consider $M$-theory
configurations including a five-brane. In both of these cases, we
derive an explicit formula for the topological invariants describing
the global anomalies.

\newsec{Anomalies on open eleven-manifolds}
We consider $M$-theory on an eleven-dimensional open spin-manifold
$Y$. The massless degrees of freedom propagating in the bulk of $Y$
are those of the eleven-dimensional supergravity multiplet, i.e. a
metric $G_{M N}$, a three-form potential $C_{M N P}$ and a fermionic
Rarita-Schwinger field $\psi_M$. At low energies, the dynamics of
these fields is governed by the eleven-dimensional supergravity
action \ref\CremmerJuliaScherk{
E. Cremmer, B. Julia and J. Scherk, `Supergravity theory in $11$
dimensions', {\it Phys. Lett.} {\bf B76} (1978) 409.
}. The bulk action possesses a classical invariance under
diffeomorphisms of $Y$, and since we are in an odd number of
space-time dimensions, this symmetry is obviously not spoiled by any
chiral anomaly.

However, there is potentially an anomaly, often called the parity
anomaly, in the bulk of $Y$, which is associated with the
Rarita-Schwinger field $\psi_M$ \ref\Wittenfour{
E. Witten, `On flux quantization in $M$-theory and the effective
action', {\tt hep-th/9609122}.
}. The operator in the kinetic term of this field is Hermitian in
eleven dimensions, but has infinitely many positive and negative
eigenvalues, leading to a potential sign problem in the definition of
the fermionic path-integral measure \AlvarezGaumeWitten. This will
show up as a change $\Delta \Gamma_{bulk}$ of the effective bulk
action $\Gamma_{bulk}$ under a diffeomorphism $\pi$ of $Y$. To
describe this anomaly, it is convenient to introduce a
twelve-dimensional manifold $(Y \times S^1)_\pi$, called the mapping
torus, as follows: We start with the cylinder $Y \times I$, where $I$
is an interval, and equip it with a metric that smoothly interpolates
between the original metric on $Y$ at one of the boundaries of $I$
and the metric obtained from it by the transformation $\pi$ at the
other boundary. Finally, we glue together the two boundaries of $I$
to form $(Y \times S^1)_\pi$. The bulk anomaly is then given by
\ref\AlvarezGaumeDellaPietraMoore{
L. Alvarez-Gaum\'e, S. Della Pietra and G. Moore, `Anomalies and odd
dimensions', {\it Ann. Phys.} {\bf 163} (1985) 288.
}
\eqn\bulkanomaly{
\Delta \Gamma_{bulk} = \pi i \left( {1 \over 2} {\rm Index}_{(Y
\times S^1)_\pi} (RS) - {3 \over 2} {\rm Index}_{(Y \times S^1)_\pi}
(D_0) \right) ,
}
where ${\rm Index}_{(Y \times S^1)_\pi} (RS)$ and ${\rm Index}_{(Y
\times S^1)_\pi} (D_0)$ denote the indices of the Rarita-Schwinger
and Dirac operators on $(Y \times S^1)_\pi$. It follows from charge
conjugation symmetry that these indices are even in twelve
dimensions, so $\Delta \Gamma_{bulk}$ is a multiple of $\pi i$,
corresponding to the sign ambiguity in the fermionic path integral
measure. Precisely in twelve dimensions, the combination of indices
that appears in \bulkanomaly\ is related to the signature $\sigma_{(Y
\times S^1)_\pi}$ of $(Y \times S^1)_\pi$ as
\eqn\miraculous{
{1 \over 8} \sigma_{(Y \times S^1)_\pi} = {\rm Index}_{(Y \times
S^1)_\pi} (RS) - 3 \, {\rm Index}_{(Y \times S^1)_\pi} (D_0) .
}
It follows that $\sigma_{(Y \times S^1)_\pi}$ is a multiple of $16$
and that the bulk anomaly can be written as
\eqn\bulkanomalytwo{
\Delta \Gamma_{bulk} = {\pi i \over 16} \sigma_{(Y \times S^1)_\pi} .
}

The massless degrees of freedom on the boundary $M$ of $Y$ include a
left-handed Rarita-Schwinger field $\psi_\mu$ and a right-handed
spinor field $\lambda$ originating from the Rarita-Schwinger field
$\psi_M$ of the bulk theory. There is also a set of left-handed
spinor fields $\chi$ in the adjoint representation of $E_8$ that
together with a set of gauge fields $A_\mu$ make up a super
Yang-Mills multiplet propagating on the boundary. These fields give
rise to an anomaly under diffeomorphisms of $Y$ that induce
diffeomorphisms of $M$, and also under gauge transformations of the
$E_8$ bundle $V$ over $M$. To describe such a transformation $\pi$,
we consider the mapping torus $(M \times S^1)_\pi$ and the $E_8$
bundle $V_\pi$ over it. These objects are constructed in analogy with
$(Y \times S^1)_\pi$ by identifying the boundaries of the cylinder $M
\times I$ after a twist by $\pi$. We note that $(Y \times S^1)_\pi$
is bounded by $(M \times S^1)_\pi$. The anomalous change $\Delta
\Gamma_{eff}$ under the transformation $\pi$ of the effective action
$\Gamma_{eff}$ obtained by integrating out the fermionic fields is
then given by a general formula derived in \Wittenone\ as
\eqn\chiral{
\Delta \Gamma_{eff} = {\pi i \over 2} \eta ,
}
where $\eta$ denotes a certain $\eta$-invariant on $(M \times
S^1)_\pi$.

The $\eta$-invariant on a closed manifold $C$ is defined as
\eqn\etainvariant{
\eta = \lim_{\epsilon \rightarrow 0} \sum_i{}^\prime {\rm sign}
(\lambda_i) \exp (-\epsilon | \lambda_i|) ,
}
where $i$ indexes the eigenvalues $\lambda_i$ of a certain operator
on $C$, and the sum runs over all $i$ such that $\lambda_i \neq 0$.
In general, the expression \etainvariant\ is prohibitively difficult
to evaluate. The situation is better if $C$ bounds some
twelve-manifold $B$, and the gauge bundle can be extended to a bundle
over $B$. Whether this is actually possible or not is a problem in
cobordism theory. In the situation at hand, where $C = (M \times
S^1)_\pi$, it is indeed possible since we can for example choose $B$
to be $(V \times S^1)_\pi$. In any case, if $C$ is the boundary of
$B$, the $\eta$-invariant on $C$ can be expressed in terms of a
certain operator $D$ on $B$. The operator $D$ is in fact the one that
arises in a calculation of the perturbative anomaly, i.e. the anomaly
under a transformation $\pi$ that can be continuously connected to
the identity, as we will now describe. The Atiyah-Singer index
theorem (see for example \ref\EguchiGilkeyHanson{
T. Eguchi, P. B. Gilkey and A. J. Hanson, `Gravitation, gauge
theories and differential geometry', {\it Phys. Rep.} {\bf 66} (1980)
213.
}) gives the index of $D$ on a {\it closed} twelve-manifold as the
integral of some characteristic class $I_{12}$. The perturbative
anomaly is then obtained through a descent procedure \ref\Zumino{
B. Zumino, `Chiral anomalies and differential geometry', in B. S.
DeWitt and R. Stora eds., {\it Relativity, groups and topology II},
North Holland, Amsterdam, 1984.
}: Since $I_{12}$ is closed, it can be written locally as $I_{12} = d
\omega_{11}$, where $\omega_{11}$ is the associated Chern-Simons
form. The latter form is not invariant under infinitesimal
diffeomorphisms and gauge transformations, but its variation is a
total derivative, i.e. $\delta \omega_{11} = d \alpha_{10}^1$, where
the superscript $1$ indicates that the form $\alpha_{10}^1$ is linear
in the parameter of the transformation. The perturbative anomaly is
now given by the integral of $\alpha_{10}^1$ over the space-time
manifold $M$. Returning to the case of an open manifold $B$ with
boundary $C$, the Atiyah-Patodi-Singer index theorem (see for example
\EguchiGilkeyHanson) now states that the $\eta$-invariant on $C$ is
given by
\eqn\APStheorem{
{1 \over 2} \eta = {\rm Index}_B (D) - \int_B I_{12} + \int_{C}
\omega_{11} .
}

In our case, $C = (M \times S^1)_\pi$. Recalling the definition of
this manifold as the cylinder $M \times I$ with the two boundaries
identified after a twist by $\pi$, we see that the last term in the
expression for ${1 \over 2} \eta$ does not really make sense, since
the integrand $\omega_{11}$ is in general not invariant under such a
transformation and therefore is not well-defined on $(M \times
S^1)_\pi$. This term should therefore more properly be written as an
integral over $M \times I$, so that the anomalous change of the
effective action is
\eqn\chiraltwo{
\Delta \Gamma_{eff} = \pi i \left( {\rm Index}_B (D) - \int_B I_{12}
+ \int_{M \times I} \omega_{11} \right).
}

As explained in \HoravaWitten, the anomaly from the $\psi_\mu$ and
$\lambda$ is given by half the standard anomaly $I_{Sugra} (R)$ from
these fields in type I supergravity, whereas the anomaly from $\chi$
is the standard anomaly $I_{SYM} (R, F)$ from this field in super
Yang-Mills theory with $E_8$ gauge group. The standard anomaly
formulas \ref\AlvarezGaumeGinsparg{
L. Alvarez-Gaum\'e and P. Ginsparg, `The structure of gauge and
gravitational anomalies', {\it Ann. Phys.} {\bf 161} (1985) 423.
} give
\eqn\standardanomaly{
\eqalign{
I_{Sugra} (R) = & {1 \over (2 \pi)^6} \left( - {1 \over 1296} ({\rm
tr} R^2)^3   + {7 \over 1080} {\rm tr} R^2 {\rm tr} R^4 - {31 \over
2835} {\rm tr} R^6 \right) \cr
I_{SYM} (R, F) = & {1 \over (2 \pi)^6} \left( - {1 \over 24} ({\rm
tr} F^2)^3 + {1 \over 16} ({\rm tr} F^2)^2 {\rm tr} R^2 - {5 \over
192} {\rm tr} F^2 ({\rm tr} R^2)^2 - {1 \over 48} {\rm tr} F^2 {\rm
tr} R^4 \right. \cr
& + \left. {31 \over 10368} ({\rm tr} R^2)^3 + {31 \over 4320} {\rm
tr} R^2 {\rm tr} R^4 + {31 \over 5670} {\rm tr} R^6 \right) ,
}
}
where $R$ and $F$ are the Riemann curvature and field-strength
two-forms respectively and ${\rm tr}$ for a power of $F$ denotes $1 /
30$ of the trace in the adjoint representation of $E_8$. Here we have
used the $E_8$ identities ${\rm tr} F^4 = {3 \over 10} ({\rm tr}
F^2)^3$ and ${\rm tr} F^6 = {1 \over 8} ({\rm tr} F^2)^3$. The
integrals of $I_{Sugra} (R)$ and $I_{SYM} (R, F)$ on a closed
twelve-manifold equal ${\rm Index} (RS) - 3 \, {\rm Index} (D_0)$ and
${\rm Index} (D_V)$ respectively, where $RS$, $D_0$ and $D_V$ are the
Rarita-Schwinger operator, the Dirac-operator and the Dirac operator
for fermions in the adjoint of $E_8$ respectively. Again, ${\rm
Index} (RS) - 3 \, {\rm Index} (D_0) = {1 \over 8} \sigma$, where
$\sigma$ denotes the signature in twelve dimensions. Indeed,
$I_{Sugra} (R)$ equals the Hirzebruch $L$-polynomial in this
dimension.

The characteristic class $I_{12} = {1 \over 2} I_{Sugra} (R) +
I_{SYM} (R, F)$ describing the anomaly from the fields $\psi_\mu$,
$\lambda$ and $\chi$ does not vanish in general, so the perturbative
anomalies from the fermionic fields do not cancel. However, it
factorizes as $I_{12} = I_4 \wedge I_8$, where
\eqn\factorization{
\eqalign{
I_4 & = {1 \over (2 \pi)^2} \left( {1 \over 4} {\rm tr} F^2 - {1
\over 8} {\rm tr} R^2 \right) \cr
I_8 & = {1 \over (2 \pi)^4} \left( - {1 \over 6} ({\rm tr} F^2)^2 +
{1 \over 6} {\rm tr} F^2 {\rm tr} R^2 - {1 \over 48} ({\rm tr} R^2)^2
- {1 \over 12} {\rm tr} R^4 \right).
}
}
We can thus write the anomalous change of the effective action as
\eqn\chiralthree{
\Delta \Gamma_{eff} = \pi i \left( {1 \over 16} \sigma_B + {\rm
Index}_B (D_V) - \int_B I_4 \wedge I_8 + \int_{M \times I} \omega_3
\wedge I_8 \right),
}
where $\omega_3$ is the Chern-Simons form of $I_4$, i.e. $d \omega_3
= I_4$.

The quantum anomaly \chiralthree\ can be cancelled by a generalized
Green-Schwarz mechanism involving the three-form potential $C$ of
eleven-dimensional supergravity \HoravaWitten\deAlwisone. With some
changes, the following discussion can also be adapted to the case of
$M$-theory on a ${\bf Z}_2$ orbifold, where the role of the boundary
is taken over by the orbifold fixed points. We begin by decomposing
$I_8$ as
\eqn\decomposition{
I_8 = - {8 \over 3} I_4 \wedge I_4 + I_8^\prime ,
}
where
\eqn\Ieightprime{
I_8^\prime = {1 \over (2 \pi)^4} \left( {1 \over 48} ({\rm tr} R^2)^2
- {1 \over 12} {\rm tr} R^4 \right).
}
The Green-Schwarz counterterms are now
\eqn\GSterms{
\Gamma_{GS} = \pi i \left( -{8 \over 3} \int_Y C \wedge G \wedge G +
\int_Y C \wedge I_8^\prime \right) ,
}
where $G = d C$ is the invariant four-form field strength. Note that
these terms are bulk interactions, although the anomaly to be
cancelled is supported on the boundary. The first of these terms is
the familiar `Chern-Simons' interaction of eleven-dimensional
supergravity. The existence of the second term can be inferred from a
one-loop calculation for type IIA strings \ref\VafaWitten{
C. Vafa and E. Witten, `A one loop test of string duality', {\it
Nucl. Phys.} {\bf B447} (1995) 261, {\tt hep-th/9505053}.
} lifted to eleven dimensions, or from the requirement of
perturbative anomaly cancelation on the $M$-theory five-brane
world-volume \DuffLiuMinasian. To be able to write this term, it is
crucial that $I_8^\prime$ depends only on $R$ and not on $F$, since
the latter field only propagates on the boundary $M$ and not in the
bulk of $Y$.

The change in the Green-Schwarz terms under the transformation $\pi$
is given by
\eqn\GSchange{
\Delta \Gamma_{GS} = \pi i \int_{Y \times \partial I} C \wedge \left(
-{8 \over 3} G \wedge G + I_8^\prime \right) ,
}
where the integral over $Y \times \partial I$ means the difference of
the integrals over $Y$ at the two boundary points of the interval
$I$. By using Stokes' theorem and the fact that $\partial(Y \times I)
= M \times I + Y \times \partial I$, we can rewrite this as
\eqn\GSchangetwo{
\Delta \Gamma_{GS} = - \pi i \int_{M \times I} C \wedge I_8 + \pi i
\int_{Y \times I} G \wedge \left( -{8 \over 3} G \wedge G +
I_8^\prime \right)  ,
}
where in the first term we have used \decomposition\ and the
condition that
\eqn\boundarycond{
G = I_4
}
on $M \times I$. This condition follows from the requirement that the
boundary interactions preserve half of the supersymmetry of the bulk
theory \HoravaWitten. In particular, the pullback of $I_4$ to $M
\times I$ is trivial in cohomology. We now assign an anomalous
transformation law to $C$ such that the quantity
\eqn\Hdefinition{
H = \omega_3 - C
}
is invariant on $M \times I$. This is consistent with the boundary
condition \boundarycond, which amounts to
\eqn\boundarycondtwo{
d H = 0
}
on $M \times I$. The anomalous change of the total action $\Gamma =
\Gamma_{bulk} + \Gamma_{eff} + \Gamma_{GS}$ can now be written as
\eqn\totalanomaly{
\eqalign{
\Delta \Gamma = & \pi i \left( {1 \over 16} \sigma_{(Y \times
S^1)_\pi} + {1 \over 16} \sigma_B + {\rm Index}_B (D_V) \right. \cr
& - \left. \int_B I_4 \wedge I_8 + \int_{(M \times S^1)_\pi} H \wedge
I_8 + \int_{(Y \times S^1)_\pi} G \wedge ( -{8 \over 3} G \wedge G +
I_8^\prime ) \right) .
}
}
Note that the since $H$ and $G$ are invariant, the integrands in the
last two terms are indeed well-defined on $(M \times S^1)_\pi$ and
$(Y \times S^1)_\pi$ respectively.

Before we continue, we will first verify that the expression
\totalanomaly\ does not depend on the choice of $B$. We can replace
$B$ by some other twelve-manifold $\tilde{B}$ with the same boundary
$(M \times S^1)_\pi$. The expression for $\Delta \Gamma$ then changes
by
\eqn\BBtilde{
\eqalign{
& \pi i \left( {1 \over 16} \sigma_{\tilde{B}} + {\rm
Index}_{\tilde{B}} (D_V) - {1 \over 16} \sigma_B  - {\rm Index}_B
(D_V) \right. \cr
& - \left. {1 \over 16} \sigma_{\tilde{B} \oplus (-B)}  - {\rm
Index}_{\tilde{B} \oplus (-B)} (D_V) \right).
}
}
Here $\tilde{B} \oplus (-B)$ denotes the closed twelve-manifold
constructed by gluing together $\tilde{B}$ and $B$ with opposite
orientation along their boundaries, and the last two terms originate
from the integrals
$- \int_{\tilde{B}} I_4 \wedge I_8 + \int_B I_4 \wedge I_8$. We can
now use the Novikov formula (see for example \ref\AtiyahSinger{
M. F. Atiyah and I. M. Singer, `Index of elliptic operators III',
{\it Ann. Math.} {\bf 87} (1968) 546.
})
\eqn\Novikovformula{
\sigma_{\tilde{B} \oplus (-B)} = \sigma_{\tilde{B}} - \sigma_B
}
to cancel the signature terms. Furthermore, it follows from charge
conjugation symmetry and the reality of the adjoint representation of
$E_8$ that ${\rm Index} (D_V)$ is always even in twelve dimensions,
so the expression \BBtilde\ vanishes modulo $2 \pi i$. Since the
action $\Gamma$ always appears as $\exp \Gamma$, such an ambiguity in
$\Delta \Gamma$ is harmless. To simplify the expression for the
anomaly, we can now choose $B = (Y \times S^1)_\pi$ and use the
properties that the signature $\sigma$ is a multiple of $16$ and
${\rm Index} (D_V)$ is even. It then follows from \totalanomaly\
that, modulo $2 \pi i$,
\eqn\totalmodulo{
\Delta \Gamma = \pi i \left( - \int_B I_4 \wedge I_8 + \int_{(M
\times S^1)_\pi} H \wedge I_8 + \int_{(Y \times S^1)_\pi} G \wedge (
-{8 \over 3} G \wedge G + I_8^\prime ) \right) .
}
We should also check that $\Delta \Gamma$ does not depend on the
precise form of $H$. For this to be true, we must actually require
$I_8$ to be trivial in cohomology on $(M \times S^1)_\pi$. (This is
of course automatic if the eight-dimensional cohomology group of this
space is trivial, as would be the case for example in
compactifications to $d \geq 4$ space-time dimensions.) It then
follows from \boundarycondtwo\ that the term in \totalmodulo\
involving $H$ actually vanishes so that
\eqn\totalagain{
\Delta \Gamma = \pi i \left( - \int_B I_4 \wedge I_8 + \int_{(Y
\times S^1)_\pi} G \wedge ( -{8 \over 3} G \wedge G + I_8^\prime )
\right) ,
}
again modulo $2 \pi i$.

The total anomaly \totalagain\ is in fact a topological invariant. In
particular, $\Delta \Gamma$ vanishes for a transformation $\pi$ that
can be continuously connected to the identity, i.e. the perturbative
anomalies cancel. To see this, we consider a smooth deformation of
the data. The variation of the characteristic classes $I_4$, $I_8$
and $I_8^\prime$ are total derivatives, and the same is true for the
field strength $G$, i.e. $\delta I_4 = d \Lambda_3$, $\delta I_8 = d
\Lambda_7$, $\delta I_8^\prime = d \Lambda_7^\prime$ and $\delta G =
d \Lambda_3^\prime$. To preserve the conditions \decomposition\ and
\boundarycond, the relations
\eqn\relations{
\eqalign{
\Lambda_7 & = - {16 \over 3} I_4 \wedge \Lambda_3^\prime +
\Lambda_7^\prime \cr
\Lambda_3 & = \Lambda_3^\prime
}
}
must hold modulo closed forms. It is then easy to see that the
expression \totalagain\ is invariant, again provided that $I_8$ is
trivial in cohomology on the boundary $(M \times S^1)_\pi$ of $B$ and
$(Y \times S^1)_\pi$.

The total anomaly $\Delta \Gamma$ is thus determined by the
topological classes of the various objects. We will not address the
difficult problem of determining the conditions for it to vanish. A
particularly interesting case is of course $M$-theory on an
eleven-manifold of the form $Y = X \times J$ for some closed
ten-manifold $X$ and an interval $J$, which is believed to describe
the strong-coupling limit of the $E_8 \times E_8$ heterotic string on
$X$ \HoravaWitten. In this case, the considerations in
\ref\Wittenthree{
E. Witten, `Global anomalies', in G. W. Gibbons, S. W. Hawking and P.
K. Townsend eds., {\it Supersymmetry and its applications}, Cambridge
University Press, Cambridge 1985.
}\ref\Killingback{
T. P. Killingback, `Global anomalies, string theory and spacetime
topology', {\it Class. Quant. Grav.} {\bf 5} (1988) 1169.
} concerning global anomalies for the $E_8 \times E_8$ heterotic
string can be carried over to the $M$-theory setting.

\newsec{Anomalies on the five-brane world-volume}
We consider an $M$-theory configuration including a five-brane. The
world-volume of the five-brane defines a six-manifold $W$ embedded in
the eleven-dimensional spin-manifold $Y$ on which the theory is
defined. (For simplicity, in this section we will only consider the
case when $Y$ is closed and orientable.) The normal bundle $N$ of $W$
in $Y$ is then an $SO(5)$ bundle over $W$. The massless fields of the
world-volume theory are those of a six-dimensional $N = 4$ tensor
multiplet \ref\Wittenfive{
E. Witten, `Five-branes and $M$-theory on an orbifold', {\it Nucl.
Phys.} {\bf B463} (1996) 383, {\tt hep-th/9512219}.
}, i.e. five scalars $\phi^i$, $i = 1, \ldots, 5$, a two-form $\beta$
with anti-selfdual field strength $T = d \beta$ and fermionic spinors
$\psi$ that take their values in the bundle $S$ constructed from the
normal bundle $N$ by using the spinor representation of the $SO(5)$
structure group.

The classical theory is invariant under diffeomorphisms of $Y$ that
map the five-brane world-volume $W$ to itself. (The invariance under
other diffeomorphisms is explicitly broken by the five-brane.) As in
the previous section, such a transformation $\pi$ is described by the
mapping torus $(Y \times S^1)_\pi$. The transformation $\pi$ induces
a diffeomorphism of $W$ and a gauge transformation on the bundle $S$,
which we describe by the mapping torus $(W \times S^1)_\pi$ and an
$SO(5)$ bundle $S_\pi$ over it. Obviously, $(W \times S^1)_\pi$ is a
seven-dimensional submanifold of the twelve-manifold $(Y \times
S^1)_\pi$. For the purpose of computing anomalies, we can regard the
world-volume theory as an $SO(5)$ gauge theory with fermions in the
spinor representation. This essentially amounts to replacing the
eleven-manifold $Y$ by the total space of the normal bundle $N$. As
stated above, the theory also contains a chiral two-form $\beta$ and
is coupled to non-dynamical gravity induced from the embedding of $W$
in $Y$. Because of the anti-selfduality constraint on the field
strength $T = d \beta$, there is no description in terms of a
covariant action, but this can be remedied by adding further
anomaly-free fields \AlvarezGaumeWitten. The anomalous change $\Delta
\Gamma_{eff}$ of the effective action $\Gamma_{eff}$ under the
transformation $\pi$ again follows from the general formula in
\Wittenone. We thus get $\Delta \Gamma_{eff} = {\pi i \over 2} \eta$,
where $\eta$ now is an $\eta$-invariant on $(W \times S^1)_\pi$.
Assuming that $(W \times S^1)_\pi$ bounds some eight-manifold $E$,
this can be expressed as
\eqn\effanomaly{
\Delta \Gamma_{eff} = \pi i \left( {\rm Index}_E (D) - \int_E J_8 +
\int_{W \times I} \omega_7 \right) .
}
The anomaly polynomial $J_8$ is here given by
\eqn\Ieight{
J_8 = {1 \over (2 \pi)^4} \left( {1 \over 256} ({\rm tr} F^2)^2 - {1
\over 192} {\rm tr} F^4 - {1 \over 384} {\rm tr} F^2 {\rm tr} R^2 -
{1 \over 768} ({\rm tr} R^2)^2 + {1 \over 192} {\rm tr} R^4 \right) ,
}
where ${\rm tr}$ for a power of $F$ denotes the trace in the
fundamental representation of $SO(5)$, which is related to the trace
${\rm Tr}$ for the spinor representation as ${\rm Tr} F^2 = {1 \over
2} {\rm tr} F^2$ and ${\rm Tr} F^4 = {3 \over 16} ({\rm tr} F^2)^2 -
{1 \over 4} {\rm tr} F^4$. The integral of $J_8$ over a closed
eight-manifold yields ${\rm Index} (D) = {1 \over 2} {\rm Index}
(D_S) - {1 \over 8} \sigma$, where $D_S$ is the Dirac operator for
chiral fermions with values in the bundle $S$ and $\sigma$ is the
signature. Finally, $\omega_7$ is the Chern-Simons form of $J_8$,
i.e. $d \omega_7 = J_8$.

The total anomaly on $W$ also receives a contribution from the bulk
theory on $Y$. The anomalous interaction is in fact the second term
in the Green-Schwarz interaction \GSterms. In the present context,
this term is better written as
\eqn\bulkterm{
\Gamma_{bulk} = \pi i \int_Y G \wedge \omega_7^\prime ,
}
where the Chern-Simons form $\omega_7^\prime$ obeys $d
\omega_7^\prime = I_8^\prime$ and $I_8^\prime$ was defined in
\Ieightprime. The reason is that the three-form $C$ is not globally
well-defined in the presence of the magnetically charged five-brane.
The field strength $G$ makes sense, though, and obeys
\eqn\Poincare{
d G = {1 \over 16} \delta_W ,
}
where $\delta_W$ is a representative of the Poincar\'e dual of $W$
supported in an infinitesimal neighborhood of $W$. (The factor of ${1
\over 16}$ is due to our normalization of $G$.) The anomalous change
of $\Gamma_{bulk}$ under the transformation $\pi$ is thus
\eqn\bulkchange{
\Delta \Gamma_{bulk} = \pi i \int_{Y \times \partial I} G \wedge
\omega_7^\prime = \pi i \left( \int_{Y \times I} G \wedge I_8^\prime
+ {1 \over 16} \int_{W \times I} \omega_7^\prime \right),
}
where we have used Stokes' theorem and \Poincare. The last term
involves the Chern-Simons form $\omega_7^\prime$ of $I_8^\prime$
restricted to $W$. To evaluate this term in the present context, one
should note that there is an important change in notation between
this section and the previous one: In the formula \Ieightprime\ for
$I_8^\prime$, $R$ denotes the curvature on $Y$, whereas in this
section we take $R$ to denote the induced curvature on $W$. We should
therefore rewrite \Ieightprime\ using the decomposition of the
tangent bundle of $Y$ restricted to $W$ as the direct sum of the
tangent bundle of $W$ and the normal bundle $N$. The latter bundle is
regarded as an $SO(5)$ bundle with field strength $F$. In this way we
get
\eqn\IeightprimeW{
I_8^\prime = {1 \over (2 \pi)^4} \left( {1 \over 48} ({\rm tr} F^2)^2
- {1 \over 12} {\rm tr} F^4 + {1 \over 24} {\rm tr} F^2 {\rm tr} R^2
+ {1 \over 48} ({\rm tr} R^2)^2 - {1 \over 12} {\rm tr} R^4 \right)
}
on $W$. The combined anomaly of $\Gamma_{eff} + \Gamma_{bulk}$ is
thus
\eqn\combined{
\Delta \Gamma_{eff} + \Delta \Gamma_{bulk} = \pi i \left( {1 \over 2}
{\rm Index}_E (D_S) - {1 \over 8} \sigma_E - \int_E J_8 + \int_{Y
\times I} G \wedge I_8^\prime + \int_{W \times I} \omega_7^{\prime
\prime} \right) ,
}
where $\omega_7^{\prime \prime}$ is the Chern-Simons form of
$I_8^{\prime \prime} = J_8 + {1 \over 16} I_8^\prime$, i.e. $d
\omega_7^{\prime \prime} = I_8^{\prime \prime}$. We see that
\eqn\Ieightprimeprime{
I_8^{\prime \prime} = {1 \over (2 \pi)^4} \left( {1 \over 192} ({\rm
tr} F^2)^2 - {1 \over 96} {\rm tr} F^4 \right) ,
}
which in fact equals $1 / 24$ times the second Pontrjagin class $p_2
(N)$ of the normal bundle $N$.

We will now describe a mechanism, outlined in \Wittentwo\ and further
elaborated in \deAlwistwo, to cancel the remaining perturbative
anomaly in \combined. In the following, we will use a vector sign
over a differential form to denote that it takes its values in the
normal bundle $N$. Bilinears in such forms are understood to be
multiplied via the fiber-metric on $N$. In this way, we can write the
characteristic class $I_8^{\prime \prime}$ as
\eqn\factorization{
I_8^{\prime \prime} = {1 \over 24} \vec{\chi} \wedge \vec{\chi} ,
}
where the $N$-valued four-form $\vec{\chi}$ is a bilinear in the
field strength $F$ contracted with the invariant rank five tensor of
$SO(5)$. (The field strength $F$ takes its values in the adjoint
representation of $SO(5)$, i.e. in the antisymmetric product of two
copies of $N$.) We also introduce the $N$-valued Chern-Simons
three-form $\vec{\omega}$ corresponding to $\vec{\chi}$ so that $D
\vec{\omega} = \vec{\chi}$, where $D$ denotes the $SO(5)$ covariant
exterior derivative. Although the square of $D$ does not vanish, it
follows from the Bianchi identity for $F$ that $D \vec{\chi} = 0$.
Furthermore, we introduce an $N$-valued three-form $\vec{H}$ as
follows: When restricted to $W$, the tangent bundle of $Y$ decomposes
as a direct sum of the tangent bundle of $W$ and the normal bundle
$N$. The field strength $G$, which is a section of the fourth
exterior power of the tangent bundle of $Y$, can be decomposed
accordingly. The form $\vec{H}$ is then proportional to the component
which is a three-form on $W$ with values in $N$. We must also require
that
\eqn\DH{
D \vec{H} = \vec{\chi}
}
when restricted to $W$, i.e. $\vec{\chi}$ must be covariantly exact.
This is a topological restriction for the anomaly cancelation
mechanism to work. It also fixes the normalization of $\vec{H}$. The
requisite counterterm is now
\eqn\counterterm{
\Gamma_{ct} = {1 \over 24} \int_W \vec{\omega} \wedge \vec{H} .
}
Its change under the transformation $\pi$ is
\eqn\counterchange{
\Delta \Gamma_{ct} = {1 \over 24} \int_{W \times \partial I}
\vec{\omega} \wedge \vec{H} = {1 \over 24} \int_{W \times I} \left(
\vec{\chi} \wedge \vec{H} - \vec{\omega} \wedge \vec{\chi} \right) ,
}
where we have used Stokes' theorem, the relationship of
$\vec{\omega}$ to $\vec{\chi}$, and the condition \DH. The anomalous
change of the total action $\Gamma_{total} = \Gamma_{eff} +
\Gamma_{bulk} + \Gamma_{ct}$ is thus
\eqn\totanomaly{
\Delta \Gamma_{total} = \pi i \left( {1 \over 2} {\rm Index}_E (D_S)
- {1 \over 8} \sigma_E - \int_E J_8 + \int_{(Y \times S^1)_\pi} G
\wedge I_8^\prime + {1 \over 24} \int_{(W \times S^1)_\pi} \vec{\chi}
\wedge \vec{H} \right) ,
}
where we have used the relationship between the Chern-Simons forms
$\omega_7^{\prime \prime}$ and $\vec{\omega}$ that follows from
\factorization. Note that the integrands of the last two terms are
well-defined on $(Y \times S^1)_\pi$ and $(W \times S^1)_\pi$
respectively, since the field strength $G$ (and thus also $\vec{H}$)
transforms covariantly under $\pi$.

We will now discuss some properties of the expression \totanomaly\
for the total anomaly. First of all, $\Delta \Gamma_{total}$ should
be independent modulo $2 \pi i$ of the choice of the eight-manifold
$E$ as long as it is bounded by $(W \times S^1)_\pi$. For this to be
true, we must, in addition to the eight-dimensional analogue of the
Novikov formula \Novikovformula, assume that
\eqn\Indexaddition{
{\rm Index}_{\tilde{E} \oplus (-E)} (D_S) = {\rm Index}_{\tilde{E}}
(D_S) - {\rm Index}_E (D_S)
}
modulo $4$ for any eight-manifolds $E$ and $\tilde{E}$ with common
boundary. Furthermore, $\Delta \Gamma_{total}$ is independent of the
exact form of $G$ and $\vec{H}$ as long as \Poincare\ and \DH\ are
fulfilled and $I_8^\prime$ and $\vec{\chi}$ are (covariantly) exact.
(The exactness of $I_8^\prime$ is again automatic in
compactifications to $d \geq 4$ dimensions, whereas the covariant
exactness of $\vec{\chi}$ is assured by \DH.) Finally, $\Delta
\Gamma_{total}$ is a topological invariant. Indeed, the variations of
the characteristic classes $\vec{\chi}$, $J_8$ and $I_8^\prime$ under
a smooth deformation of the data must be (covariantly) exact, i.e.
$\delta \vec{\chi} = D \vec{\lambda}$, $\delta J_8 = d \Lambda_7$ and
$\delta I_8^\prime = d \Lambda_7^\prime$. To preserve the
relationships \factorization\ and \DH, we must have $\delta \vec{H} =
\vec{\lambda}$ and
\eqn\lambdarel{
{1 \over 12} \vec{\lambda} \wedge \vec{\chi} = \Lambda_7 + {1 \over
16} \Lambda_7^\prime
}
modulo closed forms. It is then easy to see that the expression
\totanomaly\ is invariant.

The requirement that the anomaly vanish for any diffeomorphism $\pi$
of the eleven-manifold $Y$ that leaves the world-volume $W$ invariant
is a necessary restriction on a consistent $M$-theory configuration.
Obviously, a first question to settle is the correct interpretation
of the condition \DH, which entered already at the perturbative
level. Provided that this equation is fulfilled, it makes sense to
consider the expression \totanomaly\ for the global anomaly. It is
not clear to what extent its vanishing follows from already known
restrictions on $M$-theory configurations. Here we just remark that
there is no anomaly for pure gauge transformations, i.e.
transformations induced by diffeomorphisms of $Y$ that act trivially
on $W$. The reason is that since the homotopy group $\pi_6 (SO(5))$
is trivial, all pure gauge transformations can be continuously
connected to the identity. One should therefore consider
diffeomorphisms of $W$ that are not continuously connected to the
identity, possibly combined with gauge transformations. A basic case
is when $W$ is topologically a six-sphere $S^6$, in which case $(W
\times S^1)_\pi$ is actually the connected sum of one of the $27$
exotic seven-spheres and $S^6 \times S^1$ \Wittenone. In fact, if the
global anomaly does not vanish in this situation, it will not vanish
for any $W$. This follows from the fact that a diffeomorphism of
$S^6$ always has an analogue in the diffeomorphism group of an
arbitrary six-manifold.

\listrefs

\bye